\documentclass[twocolumn,showpacs,preprintnumbers,amsmath,amssymb]{revtex4}

\usepackage{graphicx}
\usepackage{dcolumn}
\usepackage{bm}

\begin{document}


\title{Fermi surface of a trapped dipolar Fermi gas}
\author{J.-N. Zhang and S. Yi}

\affiliation{Key Laboratory of Frontiers in Theoretical Physics, Institute of Theoretical Physics, Chinese Academy of Sciences, Beijing 100190, China}

\begin{abstract}
Under the framework of the semi-classical theory, we investigate the equilibrium state properties of a spin polarized dipolar Fermi gas through full numerical calculation. We show that the Fermi surfaces in both real and momentum spaces are stretched along the attractive direction of dipolar interaction. We further verify that the deformed Fermi surfaces can be well approximated by ellipsoids. In addition, the deformation parameters slightly depend on the local real and momentum space densities. We also study the interaction strength dependence of the energy and real and momentum space densities. By comparing them with variational results, we find that the ellipsoidal ansatz usually generates accurate results for weak dipolar interaction; while under strong dipolar interaction limit, notable discrepancy can be observed. Finally, we map out the stability boundary of the system.
\end{abstract}

\date{\today}
\pacs{03.75.Ss, 05.30.Fk, 31.15.xg, 34.20.Cf}

\maketitle

\section{Introduction}
The novel long-range and anisotropic character of dipole-dipole interaction has stimulated significant interest on studying dipolar Bose-Einstein condensate~\cite{dbec}. Experimentally, the first dipolar condensate was realized in Cr atoms, which possess a large magnetic dipole moment~\cite{stuh}. Utilizing Feshbach resonance to tune the scattering length to near zero, the dipolar effect was also observed in condensates of Rb~\cite{fatt} and Li~\cite{hulet} atoms. For spin polarized fermionic atoms, even though the contact interaction vanishes, the Pauli exclusion makes the magnetic dipole-dipole interaction difficult to observe. However, the experimental progress on trapping and cooling polar molecules, in particular the recent success in making high phase-space-density fermionic KRb gas~\cite{ye1,ye2}, opens up a new avenue for realizing strongly correlated states. Owing to the anisotropic dipolar interaction, we expect more novel features to be added to the quantum Fermi gases.

On the theoretical work of dipolar Fermi gases, You and Marinescu~\cite{you} first pointed out that the $p$-wave paired BCS states could be achieved for fermionic atoms inside an external dc field. Subsequently, the critical temperature of the superfluid transition and its relation to the trap geometry were investigated~\cite{bara,bara2}. Other theoretical work also includes studying the strongly correlated states in rapidly rotating trap~\cite{bara3,oste} and the possible biaxial nematic phases~\cite{freg2}.

For dipolar Fermi gas in normal phase, the equilibrium state properties~\cite{goral}, low-lying collective excitations~\cite{goral2}, and expansion dynamics~\cite{he} have
been studied based on the semi-classical theory. In those works, it was assumed that the momentum distribution is isotropic, which implicitly eliminates the possibility to explore the effect of exchange dipole-dipole interaction. Miyakawa {\it et al}. \cite{miya} then proposed an ellipsoidal variational ansatz for the phase space distribution function, which contains a parameter characterizing the deformation of the Fermi surface in momentum space. It was shown that the exchange dipolar interaction induced the deformation in momentum distribution and destabilized the dipolar Fermi gas. Following this work, the collective excitation and free expansion of a dipolar Fermi gas was also studied~\cite{sogo}.

In the present work, we investigate the ground state properties of a trapped dipolar Fermi gas in
normal phase. We justify the use of the ellipsoidal variational ansatz by numerically finding the
phase space distribution function. We show that, in both real and momentum spaces, the Fermi
surfaces can be approximated by ellipsoid as proposed by Miyakawa {\it et al.}~\cite{miya}. In
addition, we find that the deformation parameters are weakly dependent on the spatial and momentum
coordinates.

This paper is organized as follows. In Sec. \ref{form}, we introduce our model
and briefly outline the semi-classical theory for ultra-cold Fermi gas. In Sec. \ref{numer}, we present the numerical algorithm employed in this work. In Sec. \ref{resu}, we find numerically the phase space distribution function of a trapped dipolar Fermi gas, from which we study the equilibrium state properties of the system and compare them with those obtained variationally. Finally, we conclude in Sec. \ref{conc}.

\section{Formulation}\label{form}
We consider a system of $N$ spin polarized fermionic particles with permanent dipole moment $d$ at zero temperature. For simplicity, we assume that all dipoles are polarized along $z$-axis by an external field. The dipole-dipole interaction potential becomes
\begin{eqnarray}
V_d({\mathbf r}-{\mathbf r}')=c_d\frac{1-3\cos^2\theta_{{\mathbf
r}-{\mathbf r}'}}{|{\mathbf r}-{\mathbf r}'|^3},
\end{eqnarray}
where $\theta_{{\mathbf r}-{\mathbf r}'}$ is the angle between positive $z$-axis and the vector ${\mathbf r}-{\mathbf r}'$ and $c_d=\eta d^2/(4\pi\varepsilon_0)$ with $\eta$ being a parameter continuously tunable within the range $[-\frac{1}{2},1]$. The parameter $\eta$ can be realized using a fast rotating orienting field~\cite{giov}. It not only makes the sign of dipolar interaction changeable, it can also be used to completely switch off the dipolar interaction if $\varphi$ equals to 54.7$^\circ$, the `magic angle'. For spin polarized system, the contact interaction corresponding to $s$-wave scattering vanishes.

In second quantized form, the Hamiltonian of the system takes the form
\begin{eqnarray}
\hat H & = & \int d{\mathbf r}\hat\psi^{\dagger}(\mathbf{r})
\left[-\frac{\hbar^{2}\nabla^{2}}{2m}+U_{\mathrm{ho}}(\mathbf{r})\right]
\hat\psi(\mathbf{r})\nonumber \\
 &  & +\frac{1}{2}\int d{\mathbf r}d{\mathbf r}'
\hat\psi^{\dagger}(\mathbf{r})\hat\psi^{\dagger}(\mathbf{r}')
V_{d}(\mathbf{r}-\mathbf{r}')\hat\psi(\mathbf{r}')
\hat\psi(\mathbf{r}),
\end{eqnarray}
where $m$ is the mass of the molecule, $\hat\psi({\mathbf r})$ is the field operator, and $U_{\rm ho}({\mathbf r})=\frac{1}{2}m(\omega_\perp^2x^2+\omega_\perp^2y^2+\omega_z^2z^2)$ is the external trapping potential which is assumed to be axially symmetric.
In our numerical simulation, to make the volume of trapping potential to be a constant, we shall fix the geometric average of trap frequencies $\bar\omega=(\omega_\perp^2\omega_z)^{1/3}$. To characterize the shape of the trapping potential, we define the trap aspect ratio as $\lambda\equiv \omega_z/\omega_\perp$, such that the harmonic trap now becomes
\begin{eqnarray}
U_{\rm ho}({\mathbf r})=\frac{1}{2}m\bar\omega^2\lambda^{-2/3}(\rho^2+\lambda^2z^2),
\end{eqnarray}
where $\rho^2=x^2+y^2$.

Under Hartree-Fock mean field theory~\cite{hfmf}, the wave function of the system is a Slater determinant $\left|\Phi\right\rangle =c_{1}^{\dagger}c_{2}^{\dagger}\cdots c_{N}^{\dagger}\left|0\right\rangle$ with $c_i^\dag$ being the creation operator generating a molecule at orbital $\phi_i({\mathbf r})$. Using single-particle reduced density matrix $\rho(\mathbf{r},\mathbf{r}')=\left\langle \Phi\right|\hat\psi^{\dagger}(\mathbf{r}')\hat\psi(\mathbf{r})\left|\Phi\right\rangle$, the total energy $E=\langle\Phi|\hat H|\Phi\rangle$ can be expressed as
\begin{eqnarray}
E\!\!& = &\!\!\int d{\mathbf r}\left\{ \frac{\hbar^{2}}{2m}\left[\nabla_{\mathbf{r}}\cdot\nabla_{\mathbf{r'}} \rho(\mathbf{r},\mathbf{r}')\right]_{\mathbf{r}=\mathbf{r}'} +\rho(\mathbf{r},\mathbf{r})U_{\mathrm {ho}}(\mathbf{r})\right\} \nonumber\\
 &  &\!\! +\frac{1}{2}\int d{\mathbf r}d{\mathbf r}'\rho(\mathbf{r},\mathbf{r})\rho(\mathbf{r}',\mathbf{r}') V_{d}(\mathbf{r}-\mathbf{r}') \nonumber\\
 &  &\!\! -\frac{1}{2}\int d{\mathbf r}d{\mathbf r}'\rho(\mathbf{r}',\mathbf{r})\rho(\mathbf{r},\mathbf{r}') V_{d}(\mathbf{r}-\mathbf{r}'). \nonumber
\end{eqnarray}
To proceed further, we shall employ the Wigner function
\begin{eqnarray}
f\left(\mathbf{r},\mathbf{k}\right)=
\int d{\mathbf s}\, e^{-i\mathbf{k}\cdot\mathbf{s}}
\rho(\mathbf{r}+\frac{\mathbf{s}}{2},\mathbf{r}-\frac{\mathbf{s}}{2}),
\end{eqnarray}
which is the Fourier transform of the single-particle reduced density matrix. Under classical limit, the total energy can now be expressed as as functional of Wigner function
\begin{eqnarray}
E[f] \!\!&=&\!\! \frac{1}{(2\pi)^3}\int d{\mathbf r}d{\mathbf k}\left[
\frac{\hbar^2{\mathbf k}^2}{2m}+U_{\mathrm{ho}}(\mathbf{r})\right]f(\mathbf{r},\mathbf{k})\nonumber\\
&&\!\!+\frac{1}{2(2\pi)^6}\int d{\mathbf r}d{\mathbf k}
d{\mathbf r}'d{\mathbf k}'f(\mathbf{r},\mathbf{k}) f(\mathbf{r}',\mathbf{k}')V_{d}(\mathbf{r}-\mathbf{r}')\nonumber\\
&&\!\!-\frac{1}{2(2\pi)^6}\int d{\mathbf r}d{\mathbf k}
d{\mathbf k}'
f(\mathbf{r},\mathbf{k})f(\mathbf{r},\mathbf{k}')\widetilde V_{d}({\mathbf k}-{\mathbf k}'),\nonumber\\\label{efunc}
\end{eqnarray}
where $\widetilde V_d({\mathbf
k})=c_d\frac{4\pi}{3}(3\cos^2\theta_{\mathbf k}-1)$ is the Fourier
transform of the dipolar interaction with $\theta_{\mathbf k}$ being
the polar angle of ${\mathbf k}$. The two terms in first line of Eq.
(\ref{efunc}) correspond to, respectively, the kinetic and potential
energies, the second line represents the direct dipolar interaction
energy, and the last line is the exchange dipolar interaction
energy.

The Wigner function can be interpreted as the phase space distribution function, which is normalized to the total number of molecules
\begin{eqnarray}
N=\frac{1}{(2\pi)^3}\int d{\mathbf r}d{\mathbf k}f({\mathbf r},{\mathbf k}).\label{norm}
\end{eqnarray}
Integrating over ${\mathbf k}$ and ${\mathbf r}$, we obtain, respectively, the real space density
$n({\mathbf r})=[1/(2\pi)^{3}]\int d{\mathbf k}f({\mathbf r},{\mathbf k})$ and the momentum space
density $\tilde n({\mathbf k})=[1/(2\pi)^{3}]\int d{\mathbf r}f({\mathbf r},{\mathbf k})$. At zero
temperature, the phase space distribution function of Fermi gas takes the form of Heaviside step
function
\begin{eqnarray}
f({\mathbf r},{\mathbf k})=\Theta(S_F({\mathbf r},{\mathbf k})),\label{frk}
\end{eqnarray}
where  $S_F({\mathbf r},{\mathbf k})=0$ defines a closed surface which we refer to as the {\em Fermi surface in phase space}. For those phase points enclosed by the Fermi surface, we have $f({\mathbf r},{\mathbf k})=1$, otherwise, $f({\mathbf r},{\mathbf k})=0$.

For the simplest case, we may straightforwardly adopt the local density approximation which assumes that the Fermi energy only depends on the local real space density $n({\mathbf r})$, i.e.,
\begin{eqnarray}
S_F^{\rm(sph)}({\mathbf r},{\mathbf k})\equiv \left[6\pi^{2}n(\mathbf{r})\right]^{2/3}-{\mathbf k}^{2}.\label{sph}
\end{eqnarray}
Due to the spherical symmetry of the momentum space distribution, the exchange dipolar vanishes~\cite{goral}, such that the total energy becomes a functional of $n({\mathbf r})$ only
\begin{eqnarray}
E^{\rm(sph)}[n]&=&\int d{\mathbf r}\left[\frac{\hbar^2\left(6\pi^2n({\mathbf r})\right)^{5/3}}{20\pi^2m} +U_{\rm ho}({\mathbf r})n({\mathbf r})\right]\nonumber\\
&&+\frac{1}{2}\int d{\mathbf r}d{\mathbf r}'n({\mathbf r})n({\mathbf r}') V_{d}({\mathbf r}-{\mathbf r}').
\end{eqnarray}
The real space density can now be obtained either variationally or numerically by minimizing $E^{\rm (sph)}[n]$~\cite{goral,he}.

An improvement over Eq. (\ref{sph}) was made by Miyakawa {\it et al.}~\cite{miya}, who proposed an ellipsoidal ansatz in momentum space
\begin{eqnarray}
S_F^{\rm(ell)}({\mathbf r},{\mathbf k})\equiv \left[6\pi^{2}n(\mathbf{r})\right]^{2/3} -\left(\alpha^{-1}k_{\rho}^{2}+\alpha^{2}k_{z}^{2}\right),\label{ell}
\end{eqnarray}
where $k_\rho^2=k_x^2+k_y^2$ and the dimensionless parameter $\alpha$ characterizes the deformation of Fermi surface in momentum space. Clearly, the Fermi surface in momentum space forms an ellipsoid,
\begin{eqnarray}
\frac{k_\rho^2}{K_\perp^2}+\frac{k_z^2}{K_\parallel^2}=1,\label{ellipse}
\end{eqnarray}
where $K_\perp=\alpha^{1/2}(6\pi^2n)^{1/3}$ and $K_\parallel=\alpha^{-1}(6\pi^2n)^{1/3}$. Moreover, the real space density is assumed to have an inverted parabolic shape
\begin{eqnarray}
n({\mathbf r})=\frac{1}{6\pi^2}\left[(48N)^{1/3}\frac{\gamma}{\bar a^2}-\left(\frac{\gamma}{\bar a^2}\right)^2 \left(\beta^{-1}\rho^2+\beta^{2}z^{2}\right)\right]^{3/2},\nonumber\\\label{varden}
\end{eqnarray}
where $\bar a=\sqrt{\hbar/(m\bar\omega)}$ is the length of the harmonic oscillator. The dimensionless variational parameters $\beta$ and $\gamma$ correspond to, respectively, the deformation and compression in real space. The phase space distribution is then completely characterized by $\alpha$, $\beta$, and $\gamma$ which can be determined by minimizing the total energy. The immediate consequence of allowing the deformation in momentum space is that the exchange dipolar interaction energy is nonzero, which has a direct impact on the stability of the system~\cite{miya}.

Furthermore, based on the variational ansatz Eqs. (\ref{ell}) and (\ref{varden}), the Fermi surface in real space takes the form
\begin{eqnarray}
\frac{\rho^2}{R_\perp^2}+\frac{z^2}{R_\parallel^2}=1
\end{eqnarray}
where $R_\perp=\beta^{1/2}[3\tilde n/(4\pi)]^{1/3}$ and $R_\parallel=\beta^{-1}[3\tilde n/(4\pi)]^{1/3}$. It is worth pointing out that, in the noninteracting limit, the real space density takes the exact form
\begin{eqnarray}
n_0=\frac{1}{6\pi^2}\left[\frac{(48N)^{1/3}}{\bar a^2}-\left(\frac{1}{\bar a^2}\right)^2 \left(\lambda^{-2/3}\rho^2+\lambda^{4/3}z^{2}\right)\right]^{3/2}.\nonumber
\\\label{nonint}
\end{eqnarray}
Comparing it with Eq. (\ref{varden}), we see that, in the absence of interaction, the real space deformation $\beta_0=\lambda^{2/3}$ is completely determined by the trap aspect ratio.

In the present paper, we shall determine the phase space distribution function of a trapped dipolar Fermi gas using full numerical method and compare our results with the variational results using ellipsoidal ansatz. Before doing that, let us first briefly outline the numerical algorithm employed in this work.

\section{Numerical algorithm}\label{numer}
Due to the cylindrical symmetry of the system, the 6-dimensional phase space distribution function reduces to a 4-dimensional one as
\begin{eqnarray}
f({\mathbf r},{\mathbf k})\equiv f(\rho,z,k_\rho,k_z).
\end{eqnarray}
To obtain $f$, we shall numerically minimize the free energy
\begin{eqnarray}
{\cal F}[f]=E[f]-\mu N,
\end{eqnarray}
where the chemical potential $\mu$ is introduced to fix the total number of particles. The minimization is carried out using simulated annealing method.

To proceed, we introduce the 4-dimensional grids of extent $[0,R]\times[-Z,Z]\times[0,K_\rho]\times[-K_z,K_z]$ with total number of grid points $N_r\times N_z\times N_r\times N_z$. The values of $R$, $Z$, $K_\rho$, and $K_z$ are chosen such that we know for sure $f(\rho,z,k_\rho,k_z)=0$ for phase space point outside the grid extent. Since in both real and momentum space the system is symmetric with respect to $xy$-plane, the grid extent is reduced to $[0,R]\times[0,Z]\times[0,K_\rho]\times[0,K_z]$. For simplicity, we usually set $N_r=N_z$ in our numerical calculation, and typically, they are within the range of $40$ to $80$. For the simulated annealing method, we also introduce a fictitious temperature $T$ which controls the speed of convergence. Now, the simulation can be implemented straightforwardly as follows:
\begin{enumerate}
\item For an initial phase-space distribution $f$, a starting temperature $T$, and a chemical potential $\mu$, we calculate the free energy ${\cal F}[f]$.

\item To generate a new phase space distribution function $f'$, we select a target phase space point $g_i\equiv(\rho_i,z_i,k_{\rho i},k_{zi})$ and make a trial move by setting $f'(g_i)=1$ ($0$) if $f(g_i)=0$ ($1$). The free energy difference $\Delta{\cal F}={\cal F}[f']-{\cal F}[f]$ is then calculated. If $\Delta{\mathcal F}<0$, the trial move is accepted. Otherwise, we accept it with probability $e^{-\Delta{\cal F}/T}$.

\item Step 2 is repeated under the fixed temperature $T$ until the number of trial moves reaches an upper bound $N_{\rm TRI}$, or the number of accepted moves reaches another upper bound $N_{\rm ACC}$. Generally, $N_{\rm ACC}$ is much smaller than $N_{\rm TRI}$. In case either of these two conditions is satisfied, we lower the temperature $T$ by a small fraction, say $10\%$.

\item Step 2 and 3 are repeated until the free energy converges.
\end{enumerate}

To expediate the convergence of the simulation, we always choose the target grid points for trial moves at the vicinity of the Fermi surface in step 2, as for a grid point $g_i$ deep inside the Fermi surface, it is unlikely that $f(g_i)$ will change during the simulation. Accordingly, the typical values of $N_{\rm ACC}$ and $N_{\rm TRI}$ are proportional to the number of grid points on Fermi surface. Finally, we remark that the chemical potential $\mu$ remains unchanged once it is selected at the beginning of our numerical simulation. As a consequence, the total number of particles $N$ can only be calculated after we obtain the final phase space distribution function.

\section{Results}\label{resu}
\begin{figure}
\centering
\includegraphics[width=2.8in]{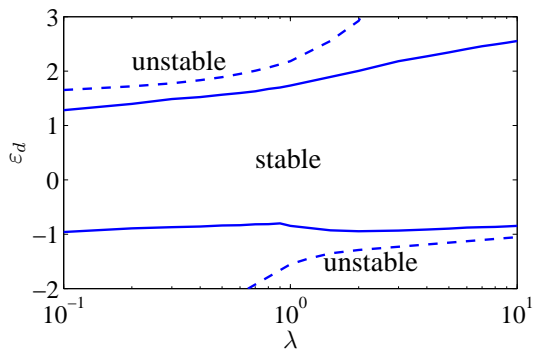}
\caption{The trap aspect ratio dependence of the critical dipolar interaction strengths from numerical calculation (solid lines). As a comparison, we also plot the variational results (dashed lines)~\cite{miya}.} \label{cri}
\end{figure}

To present our results, it is convenient to rescale all quantities into dimensionless forms. To
this end, we introduce the dimensionless units as follows: $N^{1/6}\bar a$ for length, $2\pi
N^{1/6}\bar a^{-1}$ for wavevector, and $N^{4/3}\hbar\bar\omega$ for energy. Now, the phase space
distribution function is normalized to unit, $\int d{\mathbf r}d{\mathbf k}f({\mathbf r},{\mathbf
k})=1$. The real and momentum space densities are expressed as $n({\mathbf r})=\int d{\mathbf
k}f({\mathbf r},{\mathbf k})$ and $\tilde n({\mathbf k})=[1/(2\pi)^3]\int d{\mathbf r}f({\mathbf
r},{\mathbf k})$, respectively. Here and henceforth, we adopt the same notation for the
dimensionless quantities. The dimenionless kinetic and potential energies become $2\pi^2\int
d{\mathbf r}d{\mathbf k}{\mathbf k}^2f({\mathbf r},{\mathbf k})$ and $E_{\rm pot}=\frac{1}{2}\int
d{\mathbf r}\lambda^{-2/3}(\rho^2+\lambda^2z^2)n(\mathbf{r})$, which can be calculated
straightforwardly using numerical integration. Defining a dimensionless dipolar interaction
strength
$$\varepsilon_d=\frac{N^{1/6} c_d}{\bar a^3\hbar\bar\omega},$$ the direct dipole-dipole interaction
energy can be expressed as
\begin{eqnarray}
E_{\rm dir}=\frac{\varepsilon_d}{2}\int d{\mathbf r}
d{\mathbf r}'n(\mathbf{r}) n(\mathbf{r}')\frac{1-3\cos^2\theta_{{\mathbf r}-{\mathbf r}'}}{|\mathbf{r}-\mathbf{r}'|^3},\nonumber
\end{eqnarray}
which involves only the real space density. In cylindrical coordinate, $E_{\rm dir}$ can be evaluated efficiently using Hankel transform~\cite{ronen}. Finally, the dimensionless exchange dipolar interaction energy takes the form
\begin{widetext}
\begin{eqnarray}
E_{\rm exc}
&=&\frac{2\pi\varepsilon_d}{3}\int d{\mathbf r}d{\mathbf k}d{\mathbf k}'f(\mathbf{r},\mathbf{k})f(\mathbf{r},\mathbf{k}') (1-3\cos^2\theta_{{\mathbf k}-{\mathbf k}'})\nonumber\\
&=&\frac{(2\pi)^4\varepsilon_{d}}{3}\int d\rho dzdk_{\rho}dk_{z}dk_{\rho}'dk_{z}'\rho k_{\rho}k_{\rho}'f(\rho,z,k_{\rho},k_{z}) f(\rho,z,k_{\rho}',k_{z}')\nonumber\\
&&\times\left[1-\frac{3(k_{z}-k_{z}')^{2}} {\sqrt{(k_{\rho}^{2}-k_{\rho}'^{2})^{2}+2(k_{\rho}^{2}+k_{\rho}'^{2}) (k_{z}-k_{z}')^{2}+(k_{z}-k_{z}')^{4}}}\right],\nonumber
\end{eqnarray}
\end{widetext}
where to obtain last two lines, we have analytically integrated out the azimuthal variables of ${\mathbf r}$, ${\mathbf k}$, and ${\mathbf k}'$. At first sight, this 6-dimensional integral may look formidable for numerical integration, as one can only tackle it using brutal force. Using the fact that, in each step of our numerical simulation, we only update the value of $f(\rho,z,k_\rho,k_z)$ on a single phase space point, the numerical integration can be done efficiently.

After formulating our problem into the dimensionless form, it becomes clear that the control parameters in our system reduce to the trap aspect ratio $\lambda$ and dipolar interaction strength $\varepsilon_d$. While the dependence of all physical quantities on $N$ can be easily found by converting them back to the dimensional forms~\cite{goral}. We investigate below the $\lambda$ and $\varepsilon_d$ dependences of the equilibrium properties of our system. In particular, we shall compare our results with those obtained variationally and justify the validity of the variational approach.

\subsection{Stability}
Due to the partially attractive character of dipolar interaction, there exist critical dipolar interaction strengths beyond which the system becomes unstable. This fact has been pointed out in several previous studies~\cite{goral,he,miya}. In Fig. \ref{cri}, we present the stability diagram of a dipolar Fermi gas on the $\lambda$-$\varepsilon_d$ parameter plane based on full numerical calculations. Compared to the variational results~\cite{miya}, the critical dipolar interaction strength is significantly lower. The reason that variational approach fails to predict the stability boundary is because the simple ellipsoidal ansatz, Eqs. (\ref{ell}) and (\ref{varden}), is incapable to capture the local collapse induced by strong dipolar interaction.

\begin{figure}
\centering
\includegraphics[width=2.5in]{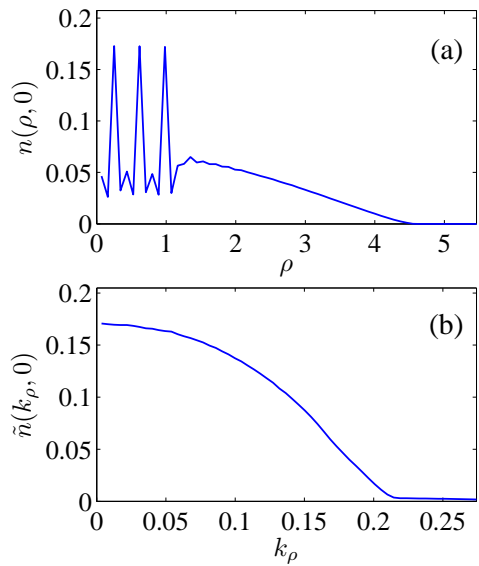}
\caption{The intermediate results for the real (a) and momentum (b) space densities of an unstable system with $\lambda=10$ and $\varepsilon_d=2.6$. The total energy diverges eventually as we continuously carry out the simulation.} \label{loclps}
\end{figure}

To illustrate the structure of local collapse, we plot in Fig. \ref{loclps} the typical intermediate real and momentum space densities from our numerical simulation. The corresponding parameters are $\lambda=10$ and $\varepsilon_d=2.6$ which fall into
the unstable region. Unlike $n({\mathbf r})$ for a stable configuration [Fig.~\ref{distall} (a) and (c)] which is a smooth function of the spatial coordinate, the real space density shown here oscillates violently. Corresponding to the sharp peaks in $n({\mathbf r})$, the momentum space density possesses a very long small valued tail, as compared to $\tilde n({\mathbf k})$ of a stable system [Fig.~\ref{distall} (b) and (d)]. We emphasize that the result shown in Fig. \ref{loclps} does not represent converged real and momentum space densities, the corresponding total energy of the system diverges eventually as we continuously carry out the simulation.

Finally, as we have properly taken into account the exchange interaction energy, which trends to destabilize the system, the critical dipolar interaction strengths presented in Fig.~\ref{cri} are also lower than those predicated numerically by using the spherical ansatz Eq. (\ref{sph})~\cite{goral,he}.

\subsection{Real and momentum space densities}
\begin{figure}
\centering
\includegraphics[width=3.4in]{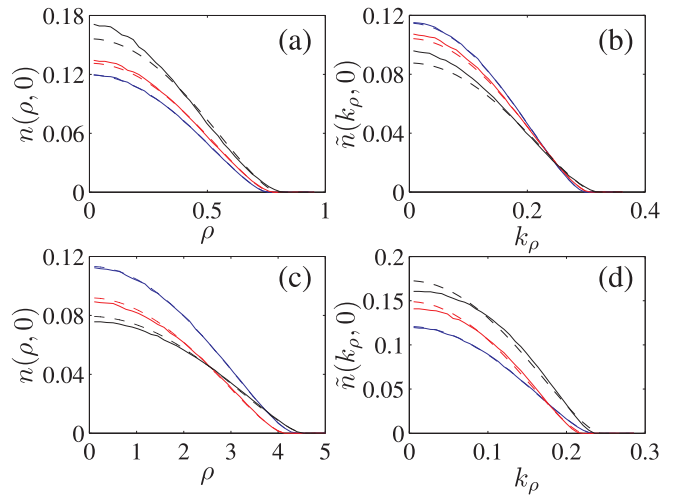}
\caption{(color online). Real and momentum space densities for $\lambda=0.1$ [(a) and (b)] and $10$ [(c) and (d)]. The solid and dashed lines correspond to, respectively, the numerical and variational results. For $\lambda=0.1$, in ascendent (descendent) order of the peak real (momentum) space density, $\varepsilon_d=0.1$, $0.5$, and $1$. For $\lambda=10$, in descendent (ascendent) order of the peak real (momentum) space density, the interaction strengths are $\varepsilon_d=0.1$, $1$, and $2$.} \label{distall}
\end{figure}

\begin{figure}
\centering
\includegraphics[width=3in]{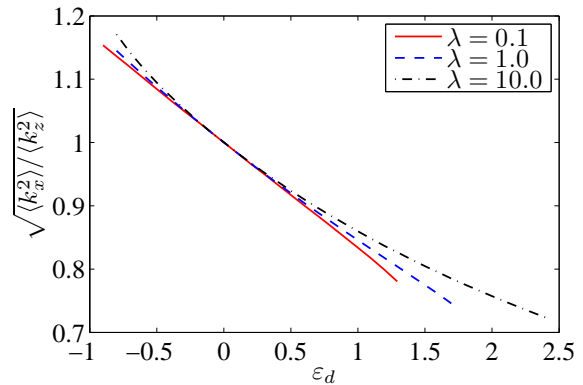}
\caption{(color online). $\sqrt{\langle k_x^2\rangle/\langle k_z^2\rangle}$ as a function of dipolar interaction strength for various trap geometries.} \label{distm}
\end{figure}

The real space density can be observed directly when the in-situ
detection is available for the system. We present, in
Fig.~\ref{distall}~(a) and (c), the typical behaviors of real space
densities corresponding to different control parameters $\lambda$
and $\varepsilon_d$. For a cigar-shaped (pancake-shaped) trap, the
peak real space density increases (decreases) as one increases the
dipolar interaction strength. This indicates that the overall
dipolar interaction is attractive (repulsive) in a cigar-shaped
(pancake-shaped) trap.  In contrast to the real space density, the
peak momentum space density [Fig. \ref{distall} (b) and (d)],
decreases (increases) as one increases the dipolar interaction
strength for a cigar-shaped (pancake-shaped) trap. As a comparison,
we also plot the real and momentum space densities obtained through
variational method. Clearly, when dipolar interaction is weak, the
agreement between two methods is very well; while under strong
dipolar interaction, the discrepancy in peak density can be as high
as 10\%.

If we switch off the dipolar interaction~\cite{giov} to let the gas expand ballistically, the momentum space density can be observed directly from the time-of-flight image. Therefore, the quantity $\kappa\equiv\sqrt{\langle k_x^2\rangle/\langle k_z^2\rangle}$ represents the deformation of the expanded cloud. We present the dipolar interaction strength dependence of $\kappa$ in Fig. \ref{distm} corresponding to different trap geometries. In noninteracting case or when the interaction is isotropic, the expanded cloud always has a spherical shape. However, the anisotropic feature of dipolar interaction always stretches the momentum space density along attractive direction of the dipolar interaction. In addition, the deformation of momentum space density only weakly depends on the trap aspect ratio.

\subsection{Energy}
\begin{figure}
\centering
\includegraphics[width=2.8in]{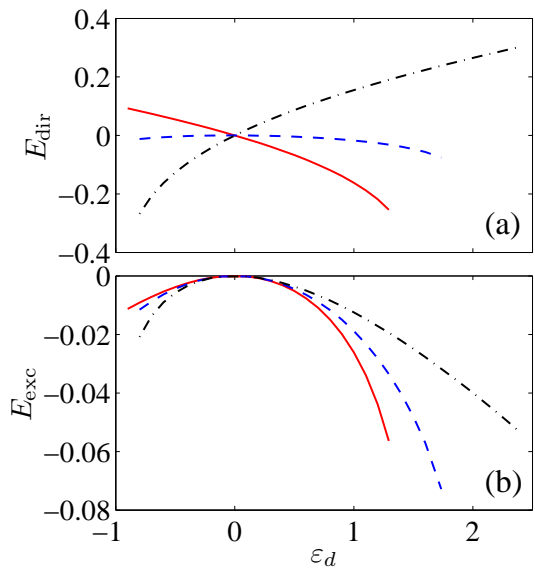}
\caption{(color online). The direct (a) and exchange (b) dipolar interaction energies for $\lambda=0.1$ (solid lines), $1$ (dashed lines), and $10$ (dash-dotted lines).} \label{edirex}
\end{figure}

In Fig. \ref{edirex} (a) and (b), we present the $\varepsilon_d$
dependence of direct and exchange interaction energies for various
trap geometries. Similar to dipolar Bose-Einstein condensate, the
direct interaction energy strongly depends on the geometry of the
trapping potential: when $\varepsilon_d>0$, $E_{\rm dir}$ is
positive (negative) for oblate (prolate) trap, indicating that the
overall direct interaction is repulsive (attractive); while in a
spherical trap, $E_{\rm dir}$ is always attractive as a result of
the stretch of the gas along the attractive direction of dipolar
interaction. On the other hand, the exchange dipolar interaction is
always negative, which reflects the fact that the momentum
distribution is always stretched along the attractive direction of
dipolar interaction. Moreover, except for in a spherical trap, where
the magnitude of $E_{\rm exc}$ is comparable to the that of $E_{\rm
dir}$, the exchange interaction energy is usually below 30\% of the
direct interaction energy in an highly anisotropic trap.

\begin{figure}
\centering
\includegraphics[width=2.8in]{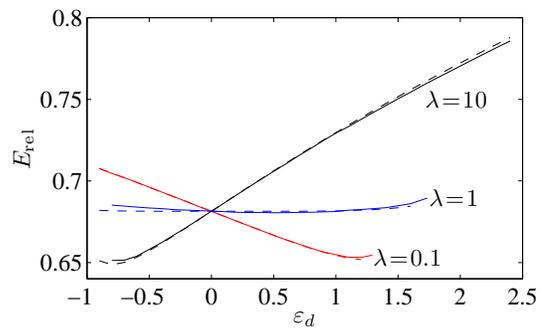}
\caption{(color online). The dipolar interaction strength dependence of the release energy for various trap aspect ratios. The solid and dashed lines correspond to, respectively, the numerical and variational results.} \label{rele}
\end{figure}

Now, we turn to study the release energy $E_{\rm rel}$ which is the sum of the kinetic and interaction energies and can be measured experimentally. In Fig.~\ref{rele}, we plot the release energy as a function of dipolar interaction strength for various trap geometries. As a result of the small exchange interaction energy, the behavior $E_{\rm rel}$ is quite similar to that obtained numerically with the spherical ansatz (\ref{sph})~\cite{he}. The dashed lines in Fig.~\ref{rele} correspond to the variational results. It can be seen that even the discrepancy increases at the strong interaction limit, it is still below 1\%. Moreover, for the total energy, the agreement between numerical and variational methods can be even better. Therefore, as long as we are in the stable region, the variational ansatz can be safely adopted to calculate the energy of the system.

\subsection{Fermi surface}
The numerical studies of the real and momentum space densities and energies of a trapped dipolar Fermi gas allows us to justify the use of variational method from the point of view of the global physical quantities. To gain more insight into the validity of variational method, we shall construct the Fermi surface of the system in phase space which enables us to compare in detail the behavior of the variational deformation parameters ($\alpha$ and $\beta$) with numerical results.

\begin{figure}
\centering
\includegraphics[width=3.2in]{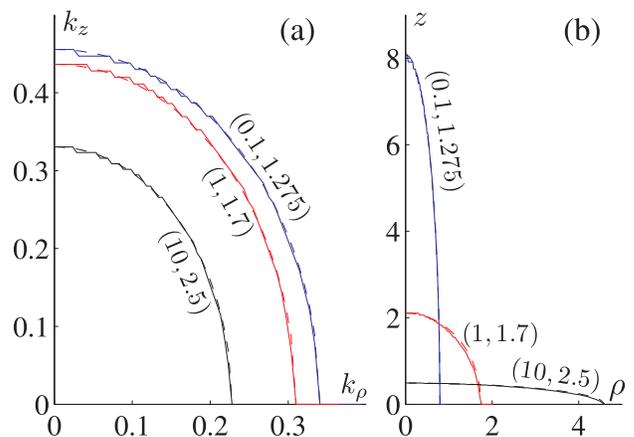}
\caption{(color online).  The Fermi surfaces in momentum space
$S_F(0,0,k_\rho,k_z)=0$ (a) and in real space $S_F(\rho,z,0,0)=0$
(b). The corresponding parameters $(\lambda,\varepsilon_d)$ are
indicated in the figures. The dashed lines in (a) and (b) are
plotted using, respectively, Eqs. (\ref{ellmome}) and
(\ref{ellreal}) with $K_{\perp,\parallel}'$ and
$R_{\perp,\parallel}'$ obtained from the numerical results.}
\label{fsall}
\end{figure}

\subsubsection{Fermi surface in momentum space}
In Fig.~\ref{fsall} (a), we plot the Fermi surfaces in momentum space, $S_F(0,0,k_\rho,k_z)=0$, corresponding to various control parameters $(\lambda,\varepsilon_d)$. Apparently, the Fermi surfaces in momentum space are anisotropic, resembling an ellipsoid described by equation
\begin{eqnarray}
\frac{k_\rho^2}{K_\perp'^2}+\frac{k_z^2}{K_\parallel'^2}=1,\label{ellmome}
\end{eqnarray}
where $K_\perp'$ and $K_\parallel'$ correspond to, respectively,
$K_\perp$ and $K_\parallel$ in the variational ansatz
Eq.~(\ref{ellipse}). By reading out the values of $K_\perp'$ and
$K_\parallel'$ from our numerical results, we plot Eq.
(\ref{ellmome}) using dashed lines in Fig.~\ref{fsall} (a). The
remarkable agreement between $S(0,0,k_\rho,k_z)=0$ and Eq.
(\ref{ellmome}) suggests that the Fermi surface in momentum space
can be well approximated by an ellipsoid. We emphasize that the
control parameters used in Fig.~\ref{fsall} fall into the strong
interaction region which are close the stability boundary (see
Fig.~\ref{cri}). For weaker dipolar interaction strength
$\varepsilon_d$, the agreement can be even better.

\begin{figure}
\centering
\includegraphics[width=3.2in]{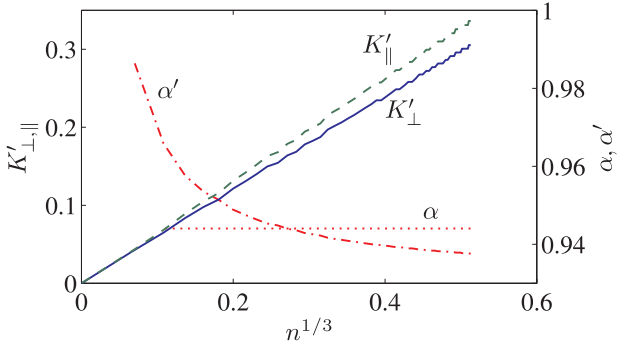}
\caption{(color online). The real density dependence of $K_\perp'$, $K_\parallel'$, and $\alpha'$ for $\lambda=0.1$ and $\varepsilon_d=0.5$. The corresponding variational $\alpha$ is plotted as dotted line.} \label{fkpp}
\end{figure}

We also find that $K_\perp'$ and $K_\parallel'$ only depend on spatial coordinate through the real space density, namely $K_{\perp,\parallel}'$ are only functions of $n({\mathbf r})$. Furthermore, as shown in Fig. \ref{fkpp}, both $K_\perp'$ and $K_\parallel'$ are roughly proportional to $n^{1/3}$, in analog to the behavior of $K_\perp$ and $K_\parallel$. To reveal more details of the deformation of Fermi surface, we define the ratio of $K_\perp$ and $K_\parallel$ as
\begin{eqnarray}
\alpha'^{3/2}\equiv\frac{K_\perp'}{K_\parallel'}.\label{kalf}
\end{eqnarray}
Apparently, $\alpha'$ corresponds to the variational parameter $\alpha$ in Eq.~({\ref{ell}) which characterizes the deformation of Fermi surface in momentum space. The typical result of $\alpha'$ is plotted in Fig. \ref{fkpp} as dash-dotted line. We immediately see that the momentum space Fermi surface is stretched along $z$-axis ($\alpha'<1$), in agreement with prediction of variational method. However, in contrast to the assumption that $\alpha$ is a constant (dotted line in Fig.~\ref{fkpp}) in the ellipsoidal ansatz, we find that $\alpha'$ is a decreasing function of $n$, which indicates that higher real space density associates with larger momentum space deformation.


\subsubsection{Fermi surface in real space}
\begin{figure}
\centering
\includegraphics[width=3.2in]{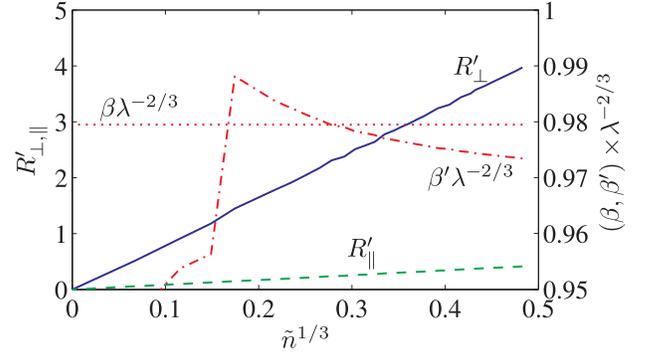}
\caption{(color online). The momentum space density dependence of $R_\perp'$, $R_\parallel'$, and $\beta'\lambda^{-2/3}$ for $\lambda=10$ and $\varepsilon_d=0.5$. The corresponding variational $\beta\lambda^{-2/3}$ is plotted as dotted line.} \label{frpp}
\end{figure}

As shown in Fig. \ref{fsall} (b), the Fermi surface in real space can be similarly expressed as an ellipsoid of the form
\begin{eqnarray}
\frac{\rho^2}{R_\perp'^2}+\frac{z^2}{R_\parallel'^2}=1.\label{ellreal}
\end{eqnarray}
In addition, the shape of $S_F(\rho,z,0,0)=0$ very sensitively depends on the trap geometries. To characterize the deformation of Fermi surface in real space, we define
\begin{eqnarray}
{\beta'}^{3/2}\equiv \frac{R_\perp'}{R_\parallel'}.
\end{eqnarray}
>From Eq. (\ref{nonint}), we immediately realize that the quantity $\beta'\lambda^{-2/3}$ represents the real space Fermi surface deformation purely induced by dipolar interaction. In Fig.~\ref{frpp}, we plot the momentum space density dependence of $R_\perp'$ and $R_\parallel'$. Similar to the momentum space case, we also find that $R_{\perp,\parallel}'$ are roughly proportional to $[\tilde n({\mathbf k})]^{1/3}$. However, the deformation parameter $\beta'$ clearly shows that higher momentum space density corresponds to larger deformation of Fermi surface in real space. In addition, the dipolar interaction alway stretches the real space Fermi surface along $z$-axis ($\beta'\lambda^{-2/3}<1$). We point out that the irregular behavior of $\beta'$ at low $\tilde n$ end is caused by the limited resolution of the grids used in numerical simulation, which becomes more important when the momentum space density is small.

%
%
%
\section{Conclusions}\label{conc}
To conclude, we have studied the equilibrium state properties of a
spin polarized dipolar Fermi gas based on the semi-classical theory.
Employing the simulated annealing method, we obtain numerically the
phase space distribution function by minimizing the total energy of
the system. We confirm that the Fermi surfaces in both real and
momentum space can be well approximated by ellipsoids, which are
stretched along the attractive direction of dipolar interaction.
However, in contrast to the ellipsoidal variational ansatz in which
the deformation parameters are assumed to be constants, we find that
they weakly depend on the local real and momentum space densities.
We also study the dipolar interaction strength dependence of the
real and momentum space densities. We find that the results from
variational calculation agree with numerical results when the
dipolar interaction is weak; while for strong dipolar interaction,
notable discrepancy is observed. Finally, we map out the stability
boundary based on numerical calculation. The numerical critical
dipolar interaction strengths are significantly lower than those
predicted variationally.

We thank Liang He for helpful discussion. This work is supported by NSFC (Grant No. 10674141), National 973 program (Grant No. 2006CB921205), and the ``Bairen" program of Chinese Academy of Sciences.

{\it Note added}. During the preparation of this manuscript, we become aware of several work on studying the ground state, sound propagation, and expansion dynamics of dipolar Fermi gases~\cite{ronen2,chan,freg,nish}. In particular, Ronen and Bohn~\cite{ronen2} obtain the exact Fermi surface of a homogeneous system through numerical calculation.

\end{document}